# Principles of Antifragile Software

Martin Monperrus

June 7, 2017

## Abstract

There are many software engineering concepts and techniques related to software errors. But is this enough? Have we already completely explored the software engineering noosphere with respect to errors and reliability? In this paper, I discuss an novel concept, called "software antifragility", that is unconventional and has the capacity to improve the way we engineer errors and dependability in a disruptive manner. This paper first discusses the foundations of software antifragilty, from classical fault tolerance to the most recent advances on automatic software repair and fault injection in production. This paper then explores the relation between the antifragility of the development process and the antifragility of the resulting software product.

## 1 Introduction

The software engineering body of knowledge on software errors and reliability is not short of concepts, starting from the classical definitions of faults, errors and failures [3], continuing with techniques for fault-freeness proofs, fault removal and fault tolerance, etc. But is this enough? Have we already completely explored the space of software engineering concepts related to errors? In this paper, I discuss a novel concept, that I call "software antifragility", which has the capacity to radically change the way we reason about software errors and the way we engineer reliability.

The notion of "antifragility" comes from the book by Nassim Nicholas Taleb simply entitled "Antifragile" [16]. Antifragility is a property of systems, whether natural or artificial: a system is antifragile if it thrives and improves when facing errors. Taleb has a broad definition of "error": it can be volatility (e.g. for financial systems), attacks and shocks (e.g. for immune systems), death (e.g. for human systems), etc. Yet, Taleb's essay is not at all about engineering, and it remains to translate the power and breadth of his vision into a set of sound engineering principles. This paper provides a first step in this direction and discusses the relations between traditional software engineering concepts and antifragility.

First, I relate software antifragility to classical fault tolerance. Second, I show the link between antifragility and the most recent advances on automatic software repair and failure injection. Third, I explore the relation between the antifragility of the development process and the antifragility of the resulting software product. This paper is a revised version of an Arxiv paper [11].

## 2 Software Antifragility

There are many pieces of evidence of software fragility, sometimes referred to as "software brittleness", [15]. For instance, the inaugural flight of Ariane 5 ended up with the total destruction of the rocket, because of an overflow in a sub-component of the system. At a totally different scale, in the Eclipse development environment, a single external plugin of a low-level optional library can crash the whole system and makes it unusable (Eclipse bug 334466). Software fragility seems independent of scale, domain and implementation technology.

There are means to combat fragility: fault prevention, fault tolerance, fault removal, and fault forecast-ing [3]. Software engineers strive for dependability. They do their best to prevent, detect and repair errors. They prevent bugs by following best practices, They detect bugs by extensively testing and comparing the implementation against the specification, They repair bugs reported by testers or users and ship the fixes in the next release. However, despite those efforts, most software remains fragile. There are pragmatic explanations to this fragility: lack of education, technical debs in legacy systems, or the economic pressure for writing cheap code. However, I think that the reason is more fundamental: we do not take the right perspective on errors.

As Taleb puts it, an antifragile system "loves errors". Software engineers do not. First, errors cost money: it is time-consuming to find and to repair bugs. Second, they are unpredictable: one can hardly forecast when and where they will occur, one can not precisely estimate the difficulty of repairing them. Software errors are traditionally considered as a plague to be eradicated and this is the problem.

Possibly, instead of damning errors, one can see them as an intrinsic characteristic of the systems we build. Complex systems have errors: in biological systems, errors constantly occur: DNA pairs are not properly copied, cells mutate, etc. Software systems of reasonable size and complexity also naturally suffer from errors, as complex biological and ecological systems do. Formal verification and model-checking fails to prove them error-free because of this very size and complexity [15]. Once one acknowledges the necessary existence of software errors in production for large and interconnected software systems [15, 12], it changes the game, it calls for new engineering principles.

## 2.1 Fault-tolerance and Antifragility

Instead of aiming at error-free software, there are software engineering techniques to constantly detect errors in production (aka self-checking software [20]) and to tolerate them as well (aka fault tolerance [13]). Self-checking, self-testing or fault-tolerance is not literally loving errors, but it is an interesting first step. Instead of shunning errors, one engineers them, or even let software crash, which is a famous motto of the Erlang community [2]. This is the right direction to go to start loving errors.

In Taleb's view, a key point of antifragility is that an antifragile system becomes better and stronger under continuous attacks and errors. The immune system, for instance, has this property: it requires constant pressure from microbes to stay reactive. Self-detection of bugs is not antifragile. Software may detect a lot of erroneous states, but it would not make it detect more.

For fault tolerance, the frontier blurs. If the fault tolerance mechanism is static there is no advantage from having more faults and there is no antifragility. If the fault tolerance mechanism is adaptive [8] and if something is learned when an error happens, the system always improves. We hit here a first characteristic of software antifragility.

> *A software system with dynamic, adaptive fault tolerance capabilities is antifragile: exposed to faults, it continuously improves.*

## 2.2 Automatic Runtime Bug Repair

Fault removal, i.e. bug repair, is one means to attain reliability [3]. Let us now consider software that repairs its own bugs at runtime and call the corresponding body of techniques "automatic runtime repair" (also called "automatic recovery" and also "self-healing" [9]).

There are two kinds of automatic software repair: state repair and behavioral repair [10]. State repair consists in modifying a program's state during its execution (the registers, the heap, the stack, etc.). Demsky and Rinard's paper on data structure repair [6] is an example of such state repair. Behavioral repair consists in modifying the program behavior, with runtime patches. The patch, whether binary or source, is synthesized and applied at runtime, with no human in the loop. For instance, the application communities of Locasto and colleagues [9] share behavioral patches for repairing faults in C code.

As said previously, a software system can be considered as antifragile as long as it learns something from bugs that occur. Automatic runtime bug repair at the behavioral level corresponds to antifragility, since each fixed bug results in a change in the code, in a better system.

> *Adaptive runtime repair means* "loving errors"*: a software system with runtime bug repair capabilities loves errors because those errors continuously trigger improvements of the system itself.*

## 2.3 Failure Injection in Production

If you really *"love errors"*, you always want more of them. In software, one can create artificial errors using techniques called fault and failure injection. So, literally, software that *"loves errors"* would continuously self-injects faults and perturbations. Would it make sense?

By self-injecting failures, a software system constantly exercises its error-recovery capabilities. If the system resists those injected failures, it will likely resist similar real-world failures. For instance, in a distributed system, servers may crash or be disconnected from the rest of the network. Consequently, a failure injector may randomly crash some servers (an example of such an injector is the Chaos Monkey [4]) to exercise the corresponding resilience capabilities.

Ensuring the occurrence of faults has three positive effects on the system. First, it forces engineers to think of error-recovery as a first-class engineering element: the system must at least be able to resist the injected faults. Second, it gives engineers and users confidence about the system's error recovery capabilities; if the system can handle those injected faults, it is likely to handle real-world natural faults of the same nature. Third, monitoring the impact of each injection gives the opportunity to learn something on the system itself and the real environmental conditions.

Because of these three effects, injecting faults in production makes the system better. This corresponds to the main characteristic of antifragility: "the antifragile loves error". It is not purely the injected faults that improve the system, it is the impact of injected faults on the engineering ecosystem (the design principles, the mindset of engineers, etc). I will come back on the profound relation between product and process in Section 3.

> A software system using fault self-injection in production is antifragile, it decreases the risk of missing, or rotting error-handling code by continuously exercising it.

Injecting faults in production must come with a careful analysis of the the dependability losses. There must be a balance between the dependability losses (due to injected system failures) and the dependability gains (due to software improvements) that result from using failure injection in production. Measuring this tradeoff is the key point of antifragile software engineering.

The idea of fault injection in production is unconventional but not new. In 1975, Yau and Cheung [20] proposed inserting fake "ghost planes" in an air traffic control system. If all the ghost planes land safely while interacting with the system and human operators, one can really trust the system. Recently, a company named Netflix released a "simian army" [7, 4], whose different kinds of monkeys inject faults in their services and datacenters. For instance, the "Chaos Monkey" randomly crashes some production servers, and the "Latency Monkey" arbitrarily increases and decreases the latency in the server network. They call this practice "chaos engineering". When fault injection is done in production on a special day under full control (as opposed to automatically at any arbitrary point in time), it is called a GameDay exercise [1].

From 1975 to today, the idea of failure injection in production has remained almost invisible. This concept is not even mentioned in the cornerstone paper by Avizienis, Laprie and Randell. [3], and the academic literature on this topic si very scarce. *However, the nascent chaos engineering community may signal a real shift.*

# 3 Software Development Process Antifragility

On the one hand, there is the software, the product, and on the other hand there is the process that builds the product. In Taleb's view, antifragility is a concept that also applies to processes. For instance, he says that the Silicon Valley innovation process is quite antifragile, because it deeply admits errors, and both inventors and investors both know that many startups will eventually fail. I now discuss the antifragility aspect of the software development process.

## 3.1 Test-driven Development

In test-driven development, developers write automated tests for each feature they write. When a bug is found, a test that reproduces the bug is first written; then the bug is fixed. The resulting strength of the test suite gives developers much confidence in the ability of their code to resist changes. Concretely, this confidence enables them to put "refactoring" as a key phase of development. Since developers have an aid (the test suite) to assess the correctness of their software, they can continuously refine the design or the implementation. They refactor fearlessly, having little doubts that they can break anything that will go unnoticed. Furthermore, test-driven development allows continuous deployment, as opposed to long release cycles. Continuous deployment means that features and bug fixes are released in production in a daily manner (and sometimes several times a day). It is the trust given by automated tests that allows continuous deployment.

What is interesting with test-driven development is the second order effect. With continuous deployment, errors have smaller impacts. No massive groups of interacting features and fixes arrive in production at the same time. When an error is found in production, the new version can be released very quickly before a catastrophic propagation.

Also, when an error is found in production, it applies to a version that is close to the most recent version of the software product (the "HEAD" version). Fixing an error in HEAD is usually much easier than fixing an error in a past version, because the patch can seamlessly be applied to all close versions, and because the developers usually have the latest version in mind. Both properties (ease of deployment, ease of fixing) contribute to minimize the effects of errors. We recognize here a property of antifragility as Taleb puts it: *"If you want to become antifragile, put yourself in the situation "loves errors" [...] by making these numerous and small in harm."* [16].

## 3.2 Bus Factor

In software development, the "bus factor" measures to what extent people are essential to a project. If a key developer is hit by a bus (or anything similar in effect), could it bring the whole project down? In dependability terms, such a consequence means that there is a failure propagation from a minor issue to a catastrophic effect.

There are management practices to cope with this critical risk. For instance, one technique is to regularly move people from projects to project, so that nobody concentrates essential knowledge. At one extreme is "If a programmer is indispensable, get rid of him as quickly as possible" [19]. In the short-term, moving people is sub-optimal. From a people perspective, they temporarily lose some productivity when they join a new project, in order to learn a new set of techniques, conventions, and communication patterns. They will often feel frustrated and unhappy because of this. From a project perspective, when a developer leaves, the project experiences a small slow-down. The slow-down lasts until the rest of the team grasps the knowledge and know-how of the developer who has just left. However, from a long-term perspective, it decreases the bus factor. In other terms, moving people transforms rare and irreversible large errors (project failure) into lots of small errors (productivity loss, slow down). This is again antifragile.

## 3.3 Conway's Law

In programming, Conway's law states that the *"organizations which design systems [...] are constrained to*

*produce designs which are copies of the communication structures of these organizations"* [5]. Raymond famously put this as *"If you have four groups working on a compiler, you'll get a 4-pass compiler"* [14]

More generally, the engineering process has an impact on the product architecture and properties. In other terms, some properties of a system emerge from the process employed to build it. Since antifragility is a property, there may be software development processes that hinder antifragility in the resulting software and others that foster it. The latter would be "antifragile software engineering".

I tend to think that the engineers that set up antifragile processes better know the nature of errors than others. I believe that developers enrolled in an antifragile process become imbued of some values of antifragility. Tseitlin's concept of "antifragile organizations" is along the same line [17]. Because of this, I hypothesize that *antifragile software development processes are better at producing antifragile software systems*.

## 4 Conclusion

This is only the beginning of antifragile software engineering. To accomplish the vision presented here, research now has to devise sound engineering techniques regarding self-checking, self-repair and failure injection in production. Because of the amount of legacy software, a major research avenue is to invent ways to develop antifragile software on top of existing brittle programming languages and execution environments. That would be a $21^{\text{th}}$ century echo to Von Neuman's dream of building reliable systems from unreliable components [18].

Acknowledgements: I would like to thank B. Cornu, M. Martinez, B. Randell, L. Seinturier, C. Vidal, E. T. Barr and the anonymous reviewers for their valuable feedback on this paper.

## References

[1] J. Allspaw. Fault injection in production. *Communications of the ACM*, 55(10):48–52, 2012.

[2] J. Armstrong. Erlang. *Communications of the ACM*, 53(9):68–75, 2010.

[3] A. Avizienis, J.-C. Laprie, B. Randell, et al. Fundamental concepts of dependability. Technical report, University of Newcastle upon Tyne, 2001.

[4] A. Basiri, N. Behnam, R. de Rooij, L. Hochstein, L. Kosewski, J. Reynolds, and C. Rosenthal. Chaos engineering. *IEEE Computer*, 33(3):35 – 41, 2016.

[5] M. E. Conway. How do committees invent? *Datamation*, 14(4):28–31, 1968.

[6] B. Demsky and M. Rinard. Automatic detection and repair of errors in data structures. *ACM SIGPLAN Notices*, 38(11):78–95, 2003.

[7] Y. Izrailevsky and A. Tseitlin. The Netflix simian army. http://techblog.netflix.com/2011/07/netflix-simian-army.html, 2011.

[8] Z. T. Kalbarczyk, R. K. Iyer, S. Bagchi, and K. Whisnant. Chameleon: A software infrastructure for adaptive fault tolerance. *IEEE Transactions on Parallel and Distributed Systems*, 10(6):560–579, 1999.

[9] M. E. Locasto, S. Sidiroglou, and A. D. Keromytis. Software self-healing using collaborative application communities. In *Proceedings of the Symposium on Network and Distributed Systems Security*, 2006.

[10] M. Monperrus. A critical review of "automatic patch generation learned from human-written patches": Essay on the problem statement and the evaluation of automatic software repair. In *Proceedings of the International Conference on Software Engineering*, 2014.

[11] M. Monperrus. Principles of antifragile software. Technical Report 1404.3056, Arxiv, 2014.

[12] H. Petroski. *To Engineer is Human: The Role of Failure in Successful Design*. Vintage Books, 1992.

[13] B. Randell. System structure for software fault tolerance. *IEEE Transactions on Software Engineering*, SE-1(2):220 –232, june 1975.

[14] E. S. Raymond et al. The jargon file. http://catb.org/jargon/, last accessed Jan. 2014, -.

[15] M. Shaw. Self-healing: softening precision to avoid brittleness. In *Proceedings of the first workshop on self-healing systems*, 2002.

[16] N. N. Taleb. *Antifragile*. Random House, 2012.

[17] A. Tseitlin. The antifragile organization. *Commun. ACM*, 56(8):40–44, Aug. 2013.

[18] J. von Neumann. Probabilistic logics and the synthesis of reliable organisms from unreliable components. *Automata Studies*, 1956.

[19] G. M. Weinberg. *The psychology of computer programming*. Van Nostrand Reinhold New York, 1971.

[20] S. Yau and R. Cheung. Design of self-checking software. In *ACM SIGPLAN Notices*, volume 10, pages 450–455. ACM, 1975.